\documentstyle{article}
\font\tenbf=cmbx10
\font\tenrm=cmr10
\font\tenit=cmti10
\font\elevenbf=cmbx10 scaled\magstep 1
\font\elevenrm=cmr10 scaled\magstep 1
\font\elevenit=cmti10 scaled\magstep 1

\renewenvironment{thebibliography}[1]
 { \elevenrm
   \begin{list}{\arabic{enumi}.}
    {\usecounter{enumi} \setlength{\parsep}{0pt}
     \setlength{\itemsep}{3pt} \settowidth{\labelwidth}{#1.}
     \sloppy
    }}{\end{list}}

\begin{document}
\begin{flushright}
{\sc IASSNS-HEP} 94-58\\{\sc cond-mat/}9408xxx, June 1994
\end{flushright}
\vspace{.3 cm}
\begin{center}{{\tenbf LIBERATING EXOTIC SLAVES\footnote{Talk at the
Celebration of
the 60th Birthday of Yakir Aharonov, February 1993.}\\}

\vglue 1.0cm
{\tenrm FRANK WILCZEK\footnote{Research supported in part by DOE
grant
DE-FG02-90ER40542} \\}
\baselineskip=13pt
{\tenit School of Natural Sciences, Institute for Advanced Study,
Olden Lane\\}
\baselineskip=12pt
{\tenit Princeton, New Jersey 08540, USA\\}
\vglue 0.8cm
{\tenrm ABSTRACT}}
\end{center}
\vglue 0.3cm
{\rightskip=3pc
 \leftskip=3pc
 \tenrm\baselineskip=12pt
 \noindent
The introduction of confined, ``slave'' fields
is frequently useful as
a formal device in
models of condensed matter physics;
it becomes a conceptual necessity for
describing possible phases of matter where the slaves are
liberated.
Here I discuss some aspects of the fractional
quantum Hall effect
from this point of view, emphasizing analogies with
phenomena in other areas of physics, particularly
to the Meissner and Higgs mechanisms,
and to confinement-deconfinement
transitions.
In this application, and in some recent attempts to model the
normal state of copper oxide superconductors, it is important
to employ slave anyon fields.
\vglue 0.6cm}

I have long admired Yakir Aharonov's style in physics: to
continue to puzzle over that which is intrinsically strange,
even in domains where more jaded spirits
have lost, from mere familiarity,  their
sense of wonder.  This child-like quality has led him to make
fundamental discoveries where few would anticipate that fundamental
discoveries could still be made, and---as we all must acknowledge
on this occasion---it obviously has kept him young!

In that spirit, I hope, I would like to discuss with you today
a personal perspective
on the fascinating complex of new states of matter forming
the
``quantum Hall complex,'' which I have developed in response
to some simple puzzles that have bothered me for a long time.
One of the puzzles, as I shall describe momentarily,
has to do with gauge invariance.  The other is broader: is the
fractional quantized Hall effect
as special and isolated as it seems at first sight, or
can its occurrence
be related to other deep ideas in theoretical physics?
I have found my perspective quite comforting and informative, and I
think it is different at least in emphasis and some
significant details from what has
appeared in the literature (including my own work.)
However, I must quickly add that
it in no way alters with
Laughlin's basic
physical picture of an incompressible quantum liquid,
nor will it be used here to derive new results that could not
be found otherwise\cite{disc,stone,fsas}.
\vglue 0.6cm
{\elevenbf\noindent 1. Critique of Laughlin's Quantization Argument}
\vglue 0.2cm
{\elevenit\noindent 1.1. The Argument}
\vglue 0.1cm
\baselineskip=14pt
\elevenrm
Shortly after the experimental discovery of the integer
quantized Hall effect, Laughlin\cite{lau1} proposed an argument, based
on gauge invariance, that explains why the conductance is
quantized.  The argument proceeds from the physical hypothesis
that in the conditions where the quantized Hall effect is observed
the electrons form an incompressible fluid in the bulk, to show
that the conductivity of the fluid (to be defined, in a precise
geometry, momentarily) must be an integer multiple of a certain
combination of fundamental constants, {\it viz}. $e^2/h$.
With some important refinements due to
Halperin\cite{halp}, this
argument remains the foundation of the theory of the effect.
I would like briefly to recall its essence.

Imagine an annulus containing electrons held at low temperature
and subject to a large perpendicular magnetic fields,
and such
that the inner and outer edges are connected by an ordinary wire
and held at a voltage difference $V$.
Suppose that we have the conditions of the quantized Hall effect,
that is, by hypothesis, that within the bulk of the annulus there
is a incompressible electron fluid.  This means that there is, for
each value of the current circulating around the annulus, a unique
bulk state of minimum energy. It can be constructed,
locally,
from the unique, isolated
ground state by a Galilean transformation.

Now let us suppose that
there is a current $I$ circulating around the annulus, and consider
the effect of switching on one quantum $h/e$ of flux in the void
within the annulus.  At the end of this operation we have produced
a gauge field, that (for electrons within the annulus)
is gauge equivalent to zero.  Thus the bulk state, assumed unique, must
return to its original form.  The only change that can have occurred,
is that some electrons from one edge might have been transferred to
the other edge, through the wire.

We can calculate the work done during this operation in two
different ways.  On the one hand, we have transferred some charge
$ne$ through a voltage $V$; thus the work is $neV$.  On the other
hand while the flux is being increased there is an azimuthal electric
field, which does work on the circulating current.  One easily
computes in this way that the work done is $(h/e) I$.  Upon equating
these, one finds for the conductance:
\begin{equation}
V/I ~=~ n e^2/h~.
\label{conductance}
\end{equation}
Thus, this transverse conductance is quantized in terms of fundamental
physical constants.


A slight variant of this argument
corresponds less well to a practical
experimental set-up, but is perhaps simpler conceptually and
will be useful for my later purposes.  Consider
the same geometry and the same process of cranking on
flux, but now with no transverse current and no voltage. As the
flux is turned on, again some integer $k$ number of electrons is
transported.  There was an azimuthal electric field as the flux
was turned on, and thus, for a determinate
transverse conductivity,
a radial current.  The electric field is proportional to the time
rate of change of the flux, so over the course of turning on one
quantum of flux there is a definite integrated radial current, or in
other words a definite charge transfer.  Equating this charge
transfer to $ke$, one finds the same quantization condition on the
transverse conductivity as before.
\vglue 0.2cm
{\elevenit\noindent 1.2. Too good to be true?}
\vglue 0.1cm
The Laughlin quantization argument is so simple and beautiful,
and so directly addresses the central phenomenon, that one
cannot
seriously doubt its essential correctness.  Unfortunately, it
is {\it too\/} good.  Shortly after it was proposed and digested,
experimentalists discovered states where the
conductance
is quantized, but now as a definite fraction of $e^2/h$ rather
than as an integer multiple.  These states occur when the
density is close to (the same) definite fraction of the density
corresponding to a full Landau level.  The jargon here is
that there is
a plateau in the resistivity around filling fraction
$\nu = \rho / (eB/\hbar c)$; meaning that when the ratio of
density to magnetic field is close to this value the conductivity
remains at the quantized value $\nu e^2/h$.
The first discovered and most robust such state
(as reflected in
the width of the associated plateau and the allowed range
of impurities and temperatures) occurs at
$\nu = 1/3$.  For simplicity and concreteness
I shall mainly focus the discussion on that state,
although by now quantized Hall
states at many other fractions have been observed and there is
a beautiful, extensive theory of them---in fact several such
theories\cite{alt}.

Now we seem to be in the embarrassing position, with the
preceding gauge invariance arguments, of having proved too
much.  The conductance is not quantized in integers
times $e^2/h$ for
incompressible bulk states, after all.   What has happened?
\vglue 0.2cm
{\elevenit\noindent 1.3. The microscopic perspective}
\vglue 0.1cm
There is a successful microscopic theory of the fractional
quantized Hall effect.  So before I get carried away with
grandiose rhetoric about breaking and
amending gauge invariance, it behooves me
to demonstrate how one  understands
at a ``mechanical'' level how the general gauge invariance
argument, which seems so clear-cut in leading to integer quantized
conductance, develops the necessary subtleties in the microscopic
theory.
\vglue 0.2cm
{\elevenit\noindent 1.4. Lightning Review of Incompressible Hall States}
\vglue 0.1cm
As we have already seen in our discussion of the
integer effect, the quantized conductance is
a fairly direct manifestation of the existence of
an incompressible quantum fluid.  That is, the
electron fluid has a preferred density
pinned to the value of the external magnetic field.  There
must be an energy gap to deviations from this
preferred density: such deviations must be accommodated by
localized inhomogeneities, rather than in arbitrarily long wavelength
``sound waves'' which---if they existed---could have arbitrarily small
energy.  In the
case of the integer quantized Hall effect the preferred density
simply corresponds to filling an integer number of Landau levels,
and the gap is quite easy to understand.  Indeed,
to raise the density
{\it here\/} and lower it {\it there\/} we must excite a particle to
the next Landau level {\it here},
which costs a finite
minimum amount of energy equal to the splitting
between Landau levels, that is not compensated
by allowing a hole {\it there}\footnote{The lowest energy
density fluctuations
actually occur at a {\it finite\/} wavevector.  These excitations,
the so-called
magnetorotons\cite{magneto}
can be regarded, intuitively, as bound states of quasiparticles and
quasiholes.  They therefore bear a family resemblance
excitons in semiconductors; however unlike most excitons
they
do not easily cascade down
and annihilate, because semiclassically
the Coulomb attraction between them---in the presence of the strong
ambient magnetic field---causes a drift in the perpendicular direction, and
thus induces
orbital motion.  Of course the magnetorotons, unlike the
quasiholes and quasiparticles discussed below, carry no net charge.}.

Laughlin himself\cite{lau2} was quick not only to recognize
the physical meaning
of the new observations, but also to propose a rationale for why
specific special (non-integer)
filling fractions should be preferred.
Let me very briefly recall the main points, since I shall
want to build on them.

First I need to remind you
of some basic results about electrons in a strong magnetic field
(here, as throughout, I am assuming that the motion of the electrons
is confined to a plane.)  The energy levels are highly degenerate
Landau levels, with a density of states $2\pi/l^2$ per unit area
per Landau level, where the magnetic length $l$ is defined
through $l^2\equiv eB/\hbar c$.  The splitting between levels
is $\hbar$ times the cyclotron frequency, {\it viz}.
$\Delta E = \hbar (eB/mc)$.
At low temperatures and for
densities small compared $2\pi l^2$ it ought to be a good
approximation to restrict attention to states formed
from single-particle states confined taken from the lowest Landau level,
unless there is some very special energetic advantage
to admixing higher levels (so as to minimize the interaction energy.)
Within the lowest Landau level, the single particle wave functions
take a particularly attractive form if one employs the so-called symmetric
gauge, defined by the vector potentials $A_x = By/2$, $A_y = -Bx/2$.
With this gauge choice, the wave functions in the lowest Landau level
take the form
\begin{equation}
\psi ~=~ f(z) e^{-{1\over 4}|z|^2 }
\label{Landaufncts}
\end{equation}
where $f(z)$ is an arbitrary analytic function of $z \equiv x + i y$,
subject to a reasonable growth condition so that the wave function
is normalizable, and distances are measured in units
of the magnetic length.  A basis of orthogonal
vectors in this Hilbert space is provided by
the functions with $f_l(z) = z^l$. $l$ is the canonical
angular
momentum around the origin, which here is intrinsically non-negative.
For reasonably large $l$, the
corresponding wave function is concentrated in a circular
ring of radius
$\sqrt {2l}$ and width $\sqrt {2\pi }$ around the origin.  It
follows, by comparing the size of the region where the wavefunction
is large to the inverse density, or by direct
calculation,
that the supports of these wave functions are highly overlapping.

Now let us consider an assembly of (non-interacting)
electrons.  Let
us suppose that they subject to a
very small potential that draws them toward the origin, but does
not appreciably change the form of the wave functions (that is
a second
order effect).  Then the ground state will
be composed out of the wave functions with the smallest values of
$l$, consistent with Fermi statistics.  It will be the
Slater determinant
\begin{equation}
\psi_1 ~=~ {\rm det} \{ z_r^{c-1} \} e^{-{1\over 4} \sum |z_k|^2}~,
\label{Slaterdet}
\end{equation}
where the row variable $r$, the column variable $c$, and $k$ all
run from 1 to $N$, the number of electrons.  Given the spatial
character of the wavefunctions as discussed above, one easily
realizes that $\psi_1$, for large values of $N$, represents a
droplet of uniform density $2\pi$ and radius $\sqrt {2N}$, with
some fuzziness in an annulus of width unity near the edge.  For later
reference let me
also record the Vandermonde identity
\begin{equation}
{\rm det} \{ z_r^{c-1} \} ~=~ \prod_{k <l: k,l = 1}^N (z_k - z_l)
\label{vandermonde}
\end{equation}

Now Laughlin's inspiration was to notice that the cube of this
wave function has remarkable qualities, that make it a particularly
attractive trial wave function for an assembly of interacting
electrons.  The Gaussian factor is then not appropriate for the
lowest Landau level, but this can be compensated by a trivial
redefinition of the length unit, which we suppose done.  Then
clearly one has a wavefunction again describing a uniform droplet
centered at the origin, now with radius $\sqrt {2N/3}$, density
$2\pi /3$ (that is, filling factor $1/3$)
and fuzziness in an annulus of width $1/\sqrt 3$ after
the rescaling.   The Laughlin wave function is particularly advantageous
if the electrons have repulsive short-range interactions, because
it enforces a triple zero as one electron approaches another.
A large number of numerical studies have shown that it is a
very good representation of the ground state wave function, for
a variety of repulsive interactions.

 From a physical point of view, the most remarkable thing about
the Laughlin wave function (and its various generalizations---see
below) is its rigidity.  It picks
out a particular filling factor in
the bulk.  Deviations from this average
density will have to be accommodated
by localized disturbances.  As we shall make much more
precise below, the situation is analogous to what one has
for type II superconductors, where magnetic fields are not
allowed in the bulk, but can penetrate only in localized
vortices.
Laughlin proposed
a form for these disturbances, that compares very well
with numerical and experimental data.  It is that a
minimal quasihole localized around $z_0$
is represented by multiplying the
wave function with a factor that pushes electrons away
from $z_0$ by adding one unit of angular momentum around
that point:
\begin{equation}
{\rm quasihole ~ factor} ~=~ \prod_1^N (z_k - z_0)~.
\label{quasifactor}
\end{equation}
This gives a density deficit; there is an analogous but
slightly more complicated construction for an enhancement, the
quasiparticle.  There is an important {\it gedanken\/} production
process for the quasihole: it is what you get by adiabatically
switching on one unit of magnetic flux at $z_0$.
The quasiholes are rather exotic: they carry fractional charge
and fractional statistics.  These properties can be shown directly
from the microscopic theory\cite{asw}.  I will forego that pleasure
here, however the result will be central to our later considerations.
\vglue 0.2cm
{\elevenit\noindent 1.5. The Gauge Argument, Reconsidered}
\vglue 0.1cm
With this background, let us return to the gauge invariance
argument.  The second form of the argument is a little easier to
discuss, so let's consider it.

There appears to be a technical
awkwardness
at the outset, in that we would like to work in an annular geometry
for the fluid and to
include some mechanism for taking electrons in one
side and out the other, whereas the simple wave functions are for
a droplet geometry.  Fortunately there is a way around this that is
quite simple and instructive for our purposes.  We have already
mentioned that
wave functions with a high power $z^l$ times the usual
exponential $e^{-{1\over 4}|z|^2}$ are concentrated in a small
ring of radius $\sqrt {2l}$ and width $\sqrt {2\pi}$
around the origin
Thus to put
a hole in the droplet of radius $R$, and produce an annulus of
quantized Hall fluid, we should multiply the wave function by
a factor
\begin{equation}
{\rm Annulizing ~ factor} ~=~ \prod_k z_k^{(R^2/2)}~.
\label{annulizer}
\end{equation}

Now you will not fail to notice that the annulizing factor is
nothing but $R^2/2$ quasiholes at the origin. A large number
of quasiholes
do literally make a (classical, spatial)
hole in the fluid!  Also, since the quasiholes are the end
result of adiabatic insertion of a unit of magnetic flux---that's
how we (following, of course, Laughlin) constructed them---we
conclude that adiabatic insertion of flux drills a hole
in the droplet.

Although
it is somewhat off the point for
this talk, it is quite interesting and appropriate to the
occasion to note that {\it by redistributing flux that
lies entirely in the empty void within the fluid annulus,
one changes the
shape of the annulus}.  Thus some of the factors
of $\prod z$ in the annulizing factor could be changed to
$\prod (z- \alpha)$.
This is a truly remarkable example of an
Aharonov-Bohm type effect, in my opinion.  That is, although
one has ``pure gauge'' outside the flux tube, by moving the
tube around one produces definite physical effects.  (There is
a pedestrian explanation for this---the moving flux tube produces an
electric field at distant points.)
The dynamics of motion within
this manifold of quasi-degenerate states, produced
by moving flux in the void, is governed by the theory of
edge excitations.  Perhaps it is even a practical proposition
to produce these
excitations by manipulating flux in this way. (End of digression.)

So now we should be able to see, in the microscopic theory, how it
can be that the gauge invariance argument becomes subtle, in such
a way that
inserting a single unit $h/e$ of flux does not transport an integral
number of electrons---while inserting three units does.

It is really
quite simple and beautiful.  The
point is that
when the power in the annulizing factor is a multiple of three, we
can again write the wavefunction in Vandermonde-Laughlin form.  That
is (stripping away the Gaussian factors):
\begin{eqnarray}
\prod_{k=1}^N z_k^{3L} \prod_{(k < l): k,l = 1}^N (z_k - z_l)^3
{}~&=& \nonumber \\
\prod_{k=1}^N z_k^{3L} ({\rm det } \{ z_r^{c - 1} \} )^3
{}~&=&\nonumber \\
({\rm det } \{ z_r^{c + L -1} \} )^3 ~,
\label{triplezero}
\end{eqnarray}
where one has $N\times N$ determinants with row index $r$ and
column index $c$.  Thus to change $L$ by one unit, to
$L+1$, we need only
to change the wavefunction of one electron, changing a $z^L$ to
a $z^{L + N}$.  In physical terms, this means removing an electron
from the inner edge and transporting it to the outer edge.  (Note
that the
{\it minimum\/}
occupied level has been emptied, and the {\it minimum\/} available
unoccupied level has been filled.)
That is
the sort of operation an ordinary wire is happy to do.  The remaining
electrons in the annular drop can be entirely passive, and need not
re-arrange their correlated wavefunctions.

It is quite a different
story if you  change the
flux by one unit.  That does not correspond to transport of an
electron from the inner edge to the outer edge, leaving the bulk
intact.  Indeed,
as we have just seen, the latter operation in its minimal
form
unambiguously corresponds
to changing the flux by {\it three\/} units.
The physical operation that corresponds to one flux unit, is
creation of a quasihole-quasiparticle pair at the
inner edge, followed by transport of the
quasiparticle to the outer edge.  This is not an operation an
ordinary wire will do for you.  There is an amplitude for it
to occur by the quasiparticle tunneling across the
sample, but since it requires a simultaneous rearrangement
of all the electrons this amplitude will be exponentially small.
In the thermodynamic limit of an infinite number of electrons, at
zero temperature, it will not occur at all.  Then we are justified in
saying that gauge invariance has been spontaneously violated, in the
only sense it ever is:
while the gauge transformation with three flux units connects
one {\it accessible\/} state to another, and represents
a legitimate symmetry; but the transformation with
a single flux unit, although formally valid, is useless
because it relates amplitudes for
processes in our world only to amplitudes for processes in
another, inaccessible one.
\vglue 0.6cm
{\elevenbf\noindent 2. Introducing, and Liberating, Confined Slaves}
\vglue 0.2cm
{\elevenit\noindent 2.1. Analogies of iQHE\footnote{I shall use this
notation for the incompressible quantum Hall effect, which
is a mouthful.  The lower case i is used here, because
IQHE is already used to indicate the integer quantized Hall effect.}
and Superconductivity}
\vglue  0.1cm
One cannot long reflect on the properties of the incompressible
Hall states without noticing many analogies between their properties
and those of ordinary superconductors.  Let me mention a few of the
most striking ones:

$\bullet$ In the quantum Hall system, there is a vanishing
longitudinal resistivity.  Thus the current flow
is non-dissipative, as in a superconductor.
Strictly speaking, this is true only
at zero temperature.  However, this fact does not spoil the analogy:
we are dealing with a two-dimensional system, and in two dimensions
the superconducting transition is also at zero temperature.  Indeed,
the reason is the same in both cases: there is a finite energy gap
to vortex production, which leads to finite though exponentially
small dissipation at any non-zero temperature.

$\bullet$ In both cases, one has an energy gap to charged
excitations.

$\bullet$ In both examples, one has {\it rigidity against
an applied magnetic field}.  In the case of superconductors
this is of course the famous
Meissner effect, but it may seem to be a rather peculiar thing
to say about iQHE states, since they occur immersed in a magnetic
field from the start.  Nevertheless they
exhibit a form of rigidity, in that changes of the field away
from a preferred value, pinned to the
effective density, are disfavored.  Here by effective density I
mean the nominal density as given by the Hall coefficient, which
is constant over a given plateau -- in the analogy, we could
call this the superfluid density.

$\bullet$ In both cases, one has vortex-like objects.  We have
of course just seen this in our discussion of the iQHE, where
the quasiparticles are in some sense vortices, and it
is a famous fact for type II superconductors.

$\bullet$ In this vein, there
is also the analogy that the non-dissipative state requires that the
vortices be pinned.  The pinning is much easier in the iQHE
case, because the vortices are electrically charged and subject to
a large magnetic field, so they will be happy to make closed orbits
on electric field equipotentials.  (Nevertheless some
impurities must be present to make these
equipotentials form closed lines,
or else there will be no plateau.  Indeed
for a translationally invariant system the Hall constant must
be equal to the carrier density, by Galilean invariance, and it
cannot ``stick'' at a preferred value as the density varies.)
At finite density the quasiparticles
would presumably, given their large
effective band mass and repulsive interactions, form a Wigner crystal,
analogous to the Abrikosov flux lattice.


On the other hand one has the apparent contrast, that the
iQHE states but not ordinary superconductors support exotic charge
and statistics
for the quasiparticles.  Also, as I discussed in the first
part of this talk, the breaking of gauge invariance is rather
different in the two cases.  For an ordinary superconductor, the
periodicity in the Aharonov-Bohm type {\it gedanken\/} experiments
we considered there would be $h/2e$ instead of the $h/(e/3)$ we
encountered for the $\nu = 1/3$ state.  The difference is profound:
whereas in the first case one has a higher degree of
flux-periodicity (that is, a smaller flux quantum) than
might of been anticipated, reflecting a pairing order
parameter, in the later case
one has a subharmonic periodicity.
\vglue 0.2cm
{\elevenit\noindent 2.2. Introducing Exotic Slaves}
\vglue 0.1 cm
The subharmonic periodicity in flux coexists, in
the iQHE,
with the existence of fractional charge, and one would
like to think that there is an organic connection between
them.  Such a connection will arise, similarly to what
one has in superconductivity, if one requires that the
integral
\begin{eqnarray}
{\rm charge~ transport~phase} ~&=&~  e^{iq \oint  A_\phi d\phi}
\nonumber \\
&=&~ e^{iq\Phi } ~,
\label{chargeint}
\end{eqnarray}
describing the phase acquired by a  particle of charge
$q$ transported
around a closed loop enclosing flux $\Phi$ to be unity,
for a {\it fractional\/} charge $q=e/3$.  This
single-valuedness, in turn, will have to be imposed
if there is condensation of a field with  charge $e/3$.
The case for an organic connection thus becomes compelling.
For the existence of fractionally  charged
quasiparticles supplies, on the face of it,
a natural candidate for
the desired condensate field: namely, of course,
the field
$\psi$
that creates the fractionally charged quasiparticles.

There is a difficulty, however.
If $\psi$ is to
condense one would like it to be
bosonic.  But that desire appears to conflict with another:
one would also like to be able to have possibility
for an
electron decay into three identical quasiparticles.  For the
quasiparticles are supposed to be the important charged low-energy
excitations, and this is the minimal decay channel that allows an
electron to communicate with them, while conserving charge.
Clearly,
if the quasiparticles are bosons this decay is not going to be possible.
One needs particles with exotic {\it anyon\/} quantum statistics,
in order that a state of three identical particles can
have the quantum numbers of a
fermion.
Furthermore the microscopic theory teaches us that the quasiparticles
are in fact anyons, and an electron can in fact decay into three
of them.
(Another possibility would have been
to have more than one kind of quasiparticle:
for example, one could reproduce the electron quantum numbers if there
were in addition a light neutral fermion excitation, so that
an electron could decay into three identical bosons and the
neutral fermion.  There may
be iQHE states with this kind of non-minimal structure---a
candidate $\nu ~=~ 1/2$ state of this kind has been
described\cite{chalf}.
However for the more conventional iQHE states, a minimalist
procedure works out quite elegantly, as we shall see.)

So we seem to have arrived at a dilemma: on the one hand
we want to have a bosonic field
to create the quasiparticles, so that the field
can condense; but on the
other hand we want the quasiparticles to be anyons, so that
they can reproduce the electron's fermion statistics.
Fortunately, these requirements  only appear to be contradictory.
Theoretical work on quantum statistics in 2+1 dimensions has
shown that a bosonic field,
properly coupled to a gauge field,
can create anyons of any type\cite{fsas}.  The way of this is done
is called
the Chern-Simons construction.  It works as follows.  One
couples the field $\psi$ using the minimal coupling procedure
to a
gauge field $a$ that does not have an ordinary Maxwell kinetic
energy term, but instead only a ``Chern-Simons'' term
\begin{equation}
\Delta {\cal L}_{\rm CS} ~=~
{n\over 4\pi } \int~ \epsilon^{\alpha \beta \gamma}
   a_\alpha f_{\beta \gamma} ~.
\label{CSL}
\end{equation}

Now one can demonstrate, without much difficulty, that
the quanta produced by $\psi$ will have their quantum statistics
altered, by the presence of the so-called
Chern-Simons gauge field $a$ of which they
are a source.
And---at least in the point particle limit, for
which the concepts are clearly defined---this change in
the statistics of the quanta is the only
effect of coupling in $a$s.  This construction
is therefore {\it a\/} valid,
and {\it the\/}
minimal, way of implementing statistical transmutation---that is,
the creation of quanta of one statistics by fields with another.

I originally called fields such
as $a$ ``fictitious'' gauge fields.
The newer terminology is in many ways preferable, but the
old terminology did have the advantage of emphasizing that the
$a$ do not introduce new local degrees of freedom.  One can
in principle fix a gauge and solve for the $a$s in terms of
$\psi$.  (The price for this is that the resulting action is
complicated and no longer manifestly local.)

Although I do not intend to pause for a full demonstration here,
it is especially
appropriate on this occasion to note that the Aharonov-Bohm
effect lies close to the heart of statistical transmutation.  For
the essence of the matter is that one finds, on solving the equations
of motion for the gauge fields $a$, that the effect of the Chern-Simons
coupling is simply to turn each quantum created by $\psi$ into
a source of flux, as well as charge.  Indeed, on varying
with respect to $a_0$ one finds the
equation
\begin{equation}
\rho~=~ -{n\over 2\pi} f_{12}
\label{fluxdensity}
\end{equation}
relating the particle number density to the Chern-Simons magnetic
field.
Note that in two space dimensions
one has flux {\it points}, as opposed to the familiar flux lines, and
one can properly speak of flux associated to a point particle.
When one such particle circles around another the wave function
acquires, as
Aharonov and Bohm taught us, a phase proportional to the product of
charge and flux.  But such a phase is operationally indistinguishable
from the effect of quantum statistics!
And that's why one can freely
change the statistics of the quanta created by a given field
$\psi$ by
coupling $\psi$ to a Chern-Simons gauge field.

We can summarize these considerations succinctly as follows.
As far as the quantum numbers of charge and statistics are
concerned, we can represent a field capable of creating an electron
as
\begin{equation}
e~\sim~\psi \psi \psi~,
\label{electronrep}
\end{equation}
where $\psi$ is a {\it bosonic\/} field with electric charge
$e/3$, properly coupled as well to a
Chern-Simons gauge field.   With our conventions, the correct
choice is simply
$n=3$ in Eq.~(\ref{CSL}).

It has frequently been useful in condensed matter problems to
introduce, as a mathematical device,
representations of electron fields as products of
other ``slave'' fields.   One might, for example, represent
the electron as a product of a neutral fermion ``spinon'' field
and
a charged boson ``holon'' field.
As long as there is a constraint in place, forbidding the separate
propagation of quanta of these fields,
this is just a mathematical device.  One is then in a confined phase,
analogous to the confined phase for quarks in QCD.
What we have done here is introduce a particular exotic kind
of slave field, with fractional charge and statistics.  As long
as its quanta are kept
confined---as might be implemented by a $Z_3$
gauge field coupling---doing this is just a mathematical device.
As long as we consider only scales much larger than the confinement
scale, we will not have changed the physical content of the theory.
The procedure will be useful if added flexibility introduced by the
slave variables allows us to represent excitations or
correlations that are
awkward to describe (i.e. non-local) in terms of the original
variables.

\vglue 0.2cm
{\elevenit\noindent 2.3. iQHE as a Modified Meissner
Effect: Liberating the Slaves}
\vglue 0.1cm
We introduced the slave field $\psi$ with two purposes in mind:
the straightforward one, that after all there are
quasiparticle states with exotic quantum numbers
in the iQHE, so we
should have fields to create them; and the deeper one, that
we would like to have a condensation, or vacuum expectation
value, of charge $e/3$ fields,
so as to understand the subharmonic
flux periodicity in the Laughlin argument.

Can $\psi$ condense? At first hearing the idea might sound
mad.  After all $\psi$ is a charged field, and the essence of
the Meissner effect is that charged fields cannot condense in
the presence of a background magnetic field.  They are, in
the jargon, frustrated.   Since the iQHE necessarily takes place
in a large background magnetic field, the proposed condensation
sounds to be grossly anti-Meissner.

On deeper consideration, however, one discovers within this
seeming difficulty the central point of this circle
of ideas.  Let us recall how one understands the Meissner effect,
in the language of condensation.  In the free energy associated
with a charged condensing field $\eta $ one has a gradient
term
\begin{equation}
|\nabla \eta |^2 ~=~ | \partial_\mu \eta - i q A_\mu \eta |^2
\label{gradterm}
\end{equation}
involving the gauge covariant derivative.   Now a constant
magnetic field introduces a vector potential $A$ which grows with
the distance, and whose effect, since it is solenoidal, cannot be
cancelled by the ordinary derivative term, which is longitudinal.
Thus to maintain a non-zero expectation value for the magnitude
of $\eta$ costs a free energy density which grows with the distance,
and this can never be favorable.

Now in the analogous considerations for our exotic slave field
$\eta$, we must include not only the electromagnetic gauge field
but also the Chern-Simons field $a$.  And then we realize, that
there is a possibility for $A$ and $a$ to cancel, thus allowing
for the possibility of a uniform condensate.
This will occur when
the part of ${e\over 3} A + a$ that grows with the distance
cancels.  That, in turn, requires that the average flux density
associated with this combination of fields vanishes.
In view of Eq.~(\ref{fluxdensity}), this occurs
when one has the relation
\begin{equation}
{e\over 3} B ~=~ b ~=~ {2\pi \over n} \rho  ~=~ \pi \rho_e~,
\label{filling}
\end{equation}
where in the third equality we have taken into account the
$n=3$ demanded by quantum statistics, and that
the quasiparticle
density is three times the electron density.  Thus the
cancellation takes place precisely at filling fraction
$\nu = 1/3$.  Whereas the ordinary Meissner effect for a
superconductor tends to exclude magnetic field, the modified
Meissner effect taking into account the statistical transmutation,
excludes
{\it deviations\/} of the magnetic field from a fixed multiple
of the density (and, of course, {\it vice versa}).   Deviations
from zero field in the superconductor, or from the desirable
density in the iQHE, are accommodated most cheaply by allowing
inhomogeneities---vortices in the first case, quasiparticles in
the second.
In fact the quasiparticles are vortices too---but in the
Chern-Simons field, not the electromagnetic field.
Only by allowing such inhomogeneities can one preserve
condensation in bulk, which requires the integrated form
of Eq.~(\ref{filling}).
That is the essence of the modified
Meissner effect.

Another feature of the situation is
that the condensation
of $\psi$ into a Higgs phase entails, as a consistency
requirement,
deconfinement of its quanta.
One cannot, after all, confine vacuum quantum numbers!
Thus the two purposes which motivated
us to introduce the confined slaves, namely on the
one hand to have fields
which described the exotic quasiparticles once they are liberated,
and on the other hand to have fields capable of condensation,
are intimately related in their realization.
\vglue 0.2cm
{\elevenit\noindent 2.4. Past and Future}
\vglue 0.1cm
Well that concludes the main story I wanted to tell you today,
and I think it is a very nice story as far as it goes.
I hope I have conveyed how
the concepts of fractional charge and statistics, the
Chern-Simons construction of the latter, and the modified Meissner
effect ineluctably come
together in a coherent account encompassing both
the iQHE and ordinary superconductivity.  It does justice,
I believe, to
the `paradoxical' nature of gauge symmetry in the fractional
quantum Hall states that one encounters upon
taking the Laughlin quantization argument
seriously, as we discussed above.

This story has both a history and, I hope, a future. I'd like briefly
to comment very briefly on these, although you should be warned that
in neither case do I speak with authority.


Girvin\cite{girvin}
stressed the analogies between superconductivity and the
iQHE very early on, made pioneering attempts to construct
a consistent, unfrustrated order parameter,
and recognized the importance of the
statistical gauge field in this regard.
Girvin and MacDonald\cite{girmac} made an important connection to the
microscopic theory.  The early
ideas were refined and extended in
important ways by Zhang, Kivelson, and Hansson\cite{zkh},
and by Read\cite{read}.
There is an interesting discussion of this body of work in
Stone's book\cite{stone}.

In previous work, as far as I know, integrally charged
condensates have been emphasized.  For example in the approach of
\cite{zkh} one couples the statistical gauge
field to the electron field to make it a
``super-fermion''---though created by a bosonic field\footnote{The
notion of ``super-fermions,'' that is of particles
for which the wave function not only changes sign---that is,
accumulates
phase $\pi$---but accumulates phase $3\pi$, say, may appear
incoherent at first sight.  After all, there is no denying
that $e^{i\pi} = e^{3i\pi}$.  However, it does have a concrete
meaning {\it operating among states within the lowest Landau level}.
For in that context the relative angular momentum must be
positive, and the effect of boosting the angular momentum by
two units is to change the spectrum of allowed values, so that
the angular momentum has to be at least three.
Without the positivity restriction on
angular momenta that operates in the lowest
Landau level the allowed spectrum would not be altered, and
the notion of ``super-fermion'' would be quite dubious.}.
This can be done with a Chern-Simons coupling $n={1\over 3}$.
With this value
the modified Meissner argument gives the same relation between
real magnetic field and electron density as was discussed above.

\bigskip

In this talk I have discussed how one is naturally led to the
fractional charge condensate.  Of course the existence of such
a condensate does not contradict the existence of an
electron condensate,  but postulates additional structure.
I think there are significant advantages to this point of view.
For example the quantization of $n$ in integers is required,
for consistency, when one considers carefully the quantization
of the Chern-Simons theory on topologically non-trivial surfaces.
The appearance of integers multiplying the Chern-Simons term, and
more generally (for iQHE states at higher levels in the
hierarchy) matrices of integers describing several coupled
Chern-Simons theories, plays a crucial role in Wen's theory of
edge states\cite{wen}.  Thus both for understanding the accuracy of the
quantization in the FQHE in a fundamental way, and for connecting
ideas about the bulk state to the successful theory of edge
states, it is important to have integers.

Having identified something like an order parameter, one might
like to continue the analogy with superconductivity by considering
inhomogeneous situations, response to external fields, and so forth,
by solving classical equations
using an effective Lagrangian, in the style of Landau and Ginzburg.
In attempting this, however, one must recognize that the
fields involved in such an effective Lagrangian cannot be
regarded as normal local 2+1 dimensional fields, because they
should only create and destroy quanta in the lowest Landau level
(which makes them effectively 1+1 dimensional).

As a concrete example,
one would like to use an effective Lagrangian to describe the
motion of quasiparticles in response to slowly varying external
electric and magnetic fields, or their scattering at small momenta.
Indeed these most basic
processes  involving quasiparticles are perhaps the
most fundamental observable processes governed
by their exotic charge
and statistics, so one would like to have an explicit description
of them.
Even in the simplest case
of the integer quantized Hall effect, where the quasiparticles are
the electrons themselves, it would seem that a more direct approach
to calculating charged
particle drifts in the lowest Landau level is
appropriate, and this has quite a different flavor from solving
simple classical field equations.
This subject needs more work\cite{lnw}.


\vglue0.2cm
{\elevenit\noindent 2.5. Coda:
Question of Statistics in Spin-Charge Separation}
\vglue 0.1cm
There are several indications that the normal state of the
CuO high temperature superconductors, for the dopings at which
they exhibit superconductivity, is an anomalous metal.
Perhaps
the most striking anomaly is the linear dependence of
resistivity on temperature, down to quite low temperatures.
This is different from what is expected for a Fermi liquid,
even after allowing for various possible complications\cite{abrahams,varma}.
On the other hand there definitely are indications that
a Fermi surface exists, at least in the sense that there is a
significant singularity
in the density of states (imaginary part of the electron
Green function) at a surface in momentum space.  However,
the size of the
Fermi surface appears in some classes of experiments, particularly
photoemission, to be roughly normal; whereas Hall effect measurements,
if interpreted as reflecting Fermi surface parameters, give a very
different picture.  Although these experiments are
not entirely straightforward to interpret (because the Fermi liquid
theory fails to describe their temperature dependence correctly,
the foundations of the analysis are insecure),
on the face of it they seem to indicate a small Fermi surface for
small doping, with positive (hole-like) carriers.  Thus they seem to
reflect not the entire electron
density, but rather its deviation from half
filling.

Motivated by these and other experimental
results, which appear to require a 2-component model,
and by experience with
1+1 dimensional models, Anderson and others have
proposed that the anomalous state is characterized by
{\it spin-charge separation}, that is the existence of separate
spin and charge degrees of freedom---spinons and holons.  Electrons
are supposed to decompose into these more basic objects.   This is
known to happen in
1+1 dimensions, even for very weak coupling\cite{oner}.
In 2+1 dimensions the situation is much less clear.   The infrared
singularities that drive 1+1 dimensional metals, even for small
coupling, to qualitatively different behaviors are substantially
weaker in 2+1 dimensions.

Nevertheless one is motivated by the
phenomenology, by the 1+1 dimensional models, and by the
``existence proof'' provided by the foregoing analysis of the
iQHE, to consider the possibility that in the CuO materials the
transition to the normal state involves a liberation of exotic
slaves.
If there are states of matter in 2+1 dimensions wherein
electrons do separate into spinons and holons, the question arises
what is the statistics of these particles.  The most obvious
assignment is boson statistics for one, fermion statistics for the
other\cite{leenag}.
On closer examination however this assignment
appears to lead to severe
difficulties\cite{leenag}.
The Bose condensation temperature tends to
be very high, and if it occurred it would lead to striking effects,
none of which are observed.   My colleagues and I suggest
instead\cite{gwz}
to consider the possibility
that both species are half-fermions.  This avoids the
Bose condensation problem.  Recent work on gauge theories\cite{gauge}
inspired by the Halperin-Lee-Read\cite{hlr} theory of the compressible
Hall states near $\nu=1/2$ suggests another advantage of assigning
fractional statistics to the spinons and holons, namely that they
lead to a pattern of anomalous behaviors at least qualitatively
suggestive of CuO phenomenology.
There is
a nominal Fermi surface, but as one approaches the Fermi momentum
there is a severe renormalization of the
effective mass, so that the singularities and temperature
dependences are not of the form
predicted by Fermi liquid theory.

A detailed account of this work will be appearing shortly.  I wanted
to mention it here as it is so closely allied to the ideas discussed
in the body of the talk, and perhaps gains some credibility from the
association.









\vglue 0.5cm
{\elevenbf\noindent References}
\vglue 0.4cm

\end{document}